\documentclass[iop,revtex4]{emulateapj}
\usepackage{lineno}

\begin{document}

\renewcommand{\topfraction}{1.0}
\renewcommand{\bottomfraction}{1.0}
\renewcommand{\textfraction}{0.0}

\title{The triple system Zeta Aquarii$^1$}

\altaffiltext{1}{Based on observations obtained  at the Southern Astrophysical Research
(SOAR) telescope. }

\author{Andrei Tokovinin}
\affil{Cerro Tololo Inter-American Observatory, Casilla 603, La Serena, Chile}
\email{atokovinin@ctio.noao.edu}

\begin{abstract}
Zeta Aquarii  is a  bright  and nearby  (28\,pc)  triple star  with a  26-year
astrometric  subsystem. Almost  a half  of the  outer  540-year visual
orbit has been  covered in 238 years of  its observations.  Both inner
and  outer  orbits  are  revised  here taking into  account  recent  direct
resolution  of the  inner  pair Aa,Ab.   The  inner orbit  has a  high
eccentricity  of  0.87   and  is  inclined  to  the   outer  orbit  by
140$\pm$10\degr,  suggesting that Kozai-Lidov  cycles take  place. The
masses of the stars Aa, B, and Ab are 1.4, 1.4, and 0.6 solar. The age
of the system  is about 3\,Gyr, and the two  main components have just
left the main sequence.  Hypothetically, this system could have formed
by a dynamical capture of the small star Ab in the twin binary Aa,B.
\end{abstract} 

\maketitle

\section{Introduction}
\label{sec:intro}

Dynamics of multiple stars  contains information on their origin. From
this  perspective, triple  systems where  elements of  both  outer and
inner  orbits  are  known  and reasonably  accurate  are  particularly
interesting,  as   relative  orbit  orientation,   period  ratio,  and
interaction      between      subsystems      can      be      studied
\citep[e.g.][]{Borkovits2016}.   However, such  well-studied multiples
are  still rare,  requiring a  substantial observational  effort. This
work deals with one of  those, $\zeta$ Aquarii.  Papers bearing nearly
the  same   title  as   the  present  one   have  been   published  by
\citet{Franz1958} and \citet{Hei1984}.

The  naked-eye  ($V=3.65$  mag)  visual  binary  $\zeta$  Aquarii  was
measured in 1779  by W.~Hershel, but an earlier  observation is listed
in the first  double-star catalog by \citet{Mayer}, so  we do not know
the name of its real discoverer.  Coordinates and basic identifiers of
the components are given  in Table~\ref{tab:ID}.  The proper motion is
$(182.9,   50.4)$\,mas~yr$^{-1}$,   the   trigonometric  parallax   is
35.50$\pm$1.26\,mas \citep{HIP2}, the  spectral type is F3III-IV.  The
two pairs  are designated  in the WDS  \citep{WDS} as  STF~2909~AB and
Ebe~1~Aa,Ab, although  they have not been discovered  by W.~Struve and
J.~Ebersberger.

\begin{figure}
\epsscale{1.0}
\plotone{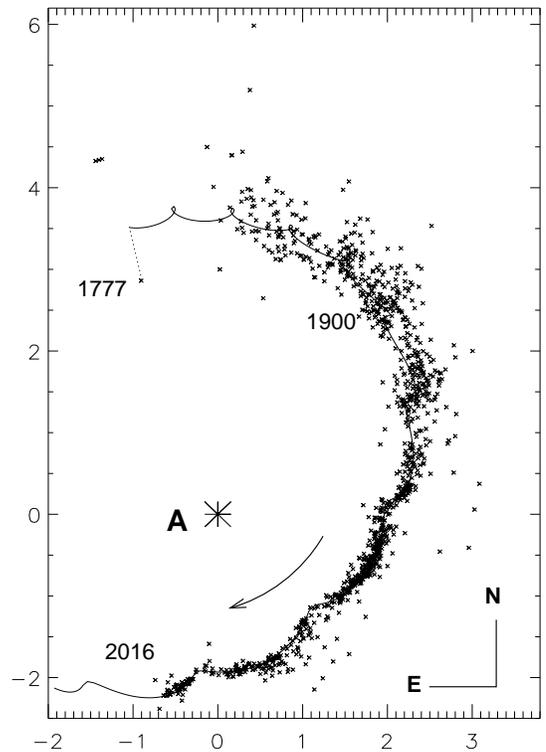}
\caption{All  observations  of  the  pair  A,B available  in  the  WDS
  database are  plotted as crosses,  together with the  proposed orbit
  (line).  The A-component  is at the coordinate origin,  the scale is in
  arcseconds. Slightly less than a  half of the outer orbit is covered
  by measures taken from 1777 to 2016.
\label{fig:hist}  }
\end{figure}

\begin{deluxetable}{l  l }[ht]
\tabletypesize{\scriptsize}
\tablewidth{0pt}
\tablecaption{Basic data on $\zeta$~A{\rm qr} \label{tab:ID}}
\tablehead{
\colhead{Object} &
\colhead{Identifiers} 
}
\startdata
AB & ADS~15971, HIP~110960, 55~Aqr \\
AB & WDS J2228$-$0001,  STF~2909~AB \\
A & 22:28:49.91  $-$00:01:11.8 (J2000) \\
A   & $\zeta_2$~Aqr, HR~8559,  HD~213052 \\
B   & $\zeta_1$~Aqr, HR~8558,  HD~213051
\enddata
\end{deluxetable}

The WDS  database contains  about 1160 measures  of the pair  A,B made
with   different    techniques   and   with    a   varying   precision
(Figure~\ref{fig:hist}).  The binary traveled  only a short arc during
the 19th century,  but in the 20th century it  became closer and moved
faster.   Photographic  measurements  available  since 1900  are  more
precise  than visual micrometer  estimates.  They  revealed deviations
from the pure Keplerian motion with a 25-year period, first discovered
by  \citet{Strand1942}.  Since then,  several authors  have undertaken
reanalysis   of   the   orbital   motions  in   this   triple   system
(Table~\ref{tab:hist}).    \citet{Harrington1968}  pointed   out  that
dynamical interaction  between components causes  measurable deviation
from  the purely  Keplerian  motion \citep[see  also  recent  work
  by][]{Xu2015}.   Yet, \citet{Hei1984} argued  that the  variation of
the osculating orbital  elements is too slow to  matter and fitted the
data by two  Keplerian orbits.  As shown below,  even this first-order
approximation of the observed motion was, so far, quite inaccurate; it
is improved here to set the stage for a future dynamical analysis.

\begin{figure}
\epsscale{0.6}
\plotone{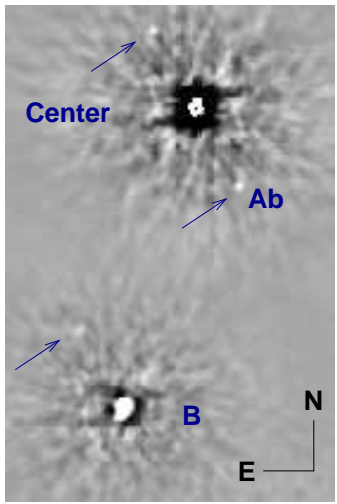}
\caption{Fragment  of  the  speckle  auto-correlation  function  (ACF)
  recorded on  2015.74 at SOAR in  the $I$ band.  The  three weak peaks
  corresponding to the tertiary  component Ab are indicated by arrows.
  The  letters mark  the positions  of Ab  and B  relative to  the ACF
  center.
\label{fig:ACF}  }
\end{figure}

The motion of visual binaries is traditionally represented by an orbit
of the secondary component B around the primary A. The ``wave'' in the
motion was  implicitly attributed  to the ``dark''  (unseen) companion
around B, until \citet{Hei1984}  demonstrated that the perturbing body
actually revolves around the  component A.  He categorically dismissed
the detections of this dark  companion to B by infrared interferometry
\citep{McCarthy1982} and  the claimed  resolution of Aa,Ab  by optical
speckle  interferometry \citep{Ebe1979},  as  they did  not match  the
orbit.   Indeed, the  large  magnitude difference  between  Aa and  Ab
($\Delta  V \approx  6$ mag)  and the  modest dynamic  range  of early
speckle instruments  made such  a resolution impossible.   Despite the
separation exceeding  0\farcs5, the  putative subsystem Aa,Ab  has not
been resolved in subsequent speckle observations until 2009. Note that
\citet{McCarthy1982} calculated  visibility from one-dimensional scans
in the North-South direction that cannot distinguish  which of the
binary  components is resolved.   Their attribution  of the
tertiary to  B was based  entirely on the  literature. At the  time of
their observation,  1981.94, the position  of Aa,Ab was  at (168\fdg6,
0\farcs23) according to  the new orbit, so it does  not contradict their
measured projected separation of 0\farcs17$\pm$0\farcs04. However, the
relative    photometry   in   \citep{McCarthy1982}    disagrees   with
our estimates, throwing doubt on this claimed resolution. 

The first real  detection of the third body  in the $\zeta$~Aqr system 
has  been made  by adaptive  optics in  the $I$  band at  the Southern
Astrophysical Research Telescope, SOAR \citep{SAM09}.  The two stars A
and B were observed separately,  and the tertiary has been erroneously
attributed  to   B.   This  has   been  corrected  in   the  following
observations, securely associating the low-mass tertiary with the main
component  A.  Direct  images  of  the subsystem  Aa,Ab  are given  by
\citet{Hrt2012}.  The tertiary Ab, 5.4  mag fainter than Aa in the $I$
band,  is  just detectable  by  modern  speckle interferometry,  which
furnishes  better  spatial  resolution  and  more  accurate  positions
compared  to  the long-exposure  imaging  with  partial AO  correction
(Figure~\ref{fig:ACF}).

The resolved  measures of  Aa,Ab do not  match the  latest astrometric
orbit of  Aa,Ab by  \citet{Scardia2010}.  Arbitrary adjustment  of the
orbit was  suggested by \citet{Tok2014}  to reach a  better agreement.
Since the work  by \citet{Hei1984}, the inner subsystem  Aa,Ab has made
one  full  revolution,  so  the  next iteration  on  the  orbits,  now
accounting for the resolved measures  of Aa,Ab, can  be made.  This
is the purpose of the present work.

\begin{deluxetable}{l   ccc ccc}
\tabletypesize{\scriptsize}
\tablewidth{0pt}
\tablecaption{History of orbit determinations of $\zeta$~A{\rm qr}  \label{tab:hist}}
\tablehead{
\colhead{Author} &
\colhead{$P_1$} & 
\colhead{$a_1$} &
\colhead{$e_1$} & 
\colhead{$P_2$} & 
\colhead{$a_2$} &
\colhead{$e_2$} \\
& \colhead{(yr)} &
\colhead{(\arcsec)} & &
\colhead{(yr)} &
\colhead{(\arcsec)} &  
}
\startdata
\citet{Strand1942}     & 400 & 3.403  & 0.60  & 25 & 0.080 & 0 \\
\citet{Franz1958}      & 600 & 4.013  & 0.45  & 25.5 & 0.097 & 0.20 \\
\citet{Harrington1968} & 856 & 5.055  & 0.50  & 25.5 & 0.072 & 0.26 \\
 \citet{Hei1984}       & 760 &  4.507 & 0.50  & 25.7 & 0.076 &  0.59 \\
Scardia+ (2010)        & 487 & 3.380  & 0.43  & 25.8 & 0.062 & 0.13 \\
This work              & 540 & 3.496  & 0.42  & 26.0 &  0.110 & 0.87 
\enddata
\end{deluxetable}

\section{The orbits}
\label{sec:orbit}

In my first attempt to update  the orbits, I used an obvious but wrong
strategy: fit the motion of A,B  in the outer orbit and then represent
the  residuals from  this motion  by  the inner  astrometric orbit  of
Aa,Ab.   The  inner  orbit   is  eccentric,  therefore  the  star  Aa,
practically  coincident with the  photo-center A,  spends most  of the
time  near  apastron and  is  displaced from  the  center  of mass  of
Aa,Ab. The outer orbit must describe the motion of B around the center
of mass,  not around the  photo-center (component A).  When  the outer
orbit is computed from the positions of A,B, the average residuals are
minimized, and the perturbation (deviations of A from the outer orbit)
becomes centrally symmetric,  leading to the false inner  orbit with a
small eccentricity. This explains  why astrometric orbits tend to have
small eccentricities  in general.   Whenever an astrometric  binary is
resolved, it often  turns out that its true visual  orbit has a larger
eccentricity  than the  astrometric orbit.   \citet{Hei1984} discussed
this  effect and insisted  on fitting  the two  orbits simultaneously.
Yet,  \citet{Scardia2010} ignored  his warning  and derived  the inner
orbit  of $\zeta$~Aqr  with  a small  eccentricity  after fitting  and
subtracting the outer orbit.

I include the  resolved measures of Aa,Ab together  with the positions
of A,B in  the data set and fit 15 free  parameters: seven elements of
A,B,  seven elements  of Aa,Ab,  and $F$,  the ratio  of the  true and
astrometric  semimajor axes  in  the inner  orbit.   If the  subsystem
belonged to  the B,  the parameter $F$  would be negative  (the reflex
displacement of  B is  oposite to  the vector Ba,Bb),  but here  it is
positive.  I modified  the  IDL code  {\tt orbit.pro}\footnote{  {\url
    http://www.ctio.noao.edu/\~{}atokovin/orbit/index.html}. See also 
\url{http://dx.doi.org/10.5281/zenodo.61119}}   to  fit
both orbits simultaneously. 

The  least-squares  fitting implicitly  assumes  that the  measurement
errors are  known and normally  distributed.  In this ideal  case, the
weights are  inversely proportional to the squares  of the measurement
errors $\sigma$.   It is well  known, however, that  visual micrometer
measures of  binary stars are  not normally distributed,  showing both
large    occasional   deviations    and    systematic   errors    (see
Figure~\ref{fig:hist}).  There is no general rule allowing to describe
these  data by  normally distributed  random  variables.  Photographic
measurements of  binary stars  can also have  larger-than-usual errors
caused  by  poor  seeing,  systematic  errors  proper  to  photography
(interaction of closely spaced images in the emulsion), and subjective
errors of persons who measured the plates.  The CCD-based measurements
should be better  than the photography, although this  is not the case
for some  low-quality CCD measures  made by amateurs.  In  the speckle
interferometry,  the  uncertainty  of  the pixel  scale  and  detector
orientation, difficult  to quantify, is  often a major  contributor to
the  position errors  of wide  pairs such  as  $\zeta$~Aqr.  Published
errors  of speckle  measures  do not  always  include the  calibration
errors.

The only  viable approach  to this  problem is to  censor the  data by
rejecting obvious  outliers and to  treat the errors of  the remaining
measurements as  unknowns, estimating  them from the  orbital solution
itself.  The  large number of measurements of  $\zeta$~Aqr favors this
strategy.   I used averaged  micrometer measures  made prior  to 1900,
taking  them from  the  paper by  \citet{Hei1984},  and assumed  their
errors  of 0\farcs1.   All  micrometer measures  made  after 1900  are
ignored,  as  more accurate  and  objective photographic  observations
became available. I adopted  the errors of photographic measures prior
to  1940 as  50 mas,  then as  30 mas,  same as  for the  accepted CCD
measures.  The speckle  measures are assumed to have errors  of 3 to 9
mas.  Obvious outliers were deleted.   Then, in the process of fitting
the orbits,  those measures that deviated  by more than  $3 \sigma$ in
either  coordinate were  given an  increased $\sigma$  to  bring their
deviation  to the  $1 \sigma$  level. After  several rounds  of manual
error  adjustment  and   automatic  down-weighting  of  outliers,  the
resulting normalized goodness of  fit $\chi^2/N$ ($N$ being the number
of degrees  of freedom) became close  to one in  both coordinates. The
322 retained  observations of A,B, their adopted  errors, and residuals
to  the orbits are  listed in  Table~\ref{tab:obsAB}. Given  the large
number of measures,  it is not practical to cite  references to all of
them.  A similar Table~\ref{tab:obsAa}  gives the resolved measures of
Aa,Ab, all coming from the  speckle interferometry at SOAR. The latest
measure  in 2016.39 is still unpublished.

In the following, I denote the orbital elements of the outer and inner
pairs by  the indices 1 and  2, respectively, and  use common notation
($P$ -- period, $T_0$ -- epoch of periastron, $e$ -- eccentricity, $a$
-- semimajor axis, $\Omega$ -- position angle of the node, $\omega$ --
argument  of periastron,  $i$  -- orbital  inclination).  The  orbital
elements   found   by   fitting    15   parameters   are   listed   in
Table~\ref{tab:orb}.  For  the inner orbit,  the astrometric semimajor
axis $a_2$  is given; it should be  multiplied by $F$ to  get the true
axis.  The weighted  rms residuals to the measures  of A,B are 15\,mas
in  both  coordinates.   The  resolved  measures  of  Aa,Ab  have  rms
residuals of  10\,mas. They are  less accurate than most  speckle data
from SOAR because of the large magnitude difference between Aa and Ab.
The  periods $P_1$  and $P_2$  are similar  to their  values  found in
the previous studies, but the inner eccentricity $e_2= 0.87$ found here is
substantially larger (see Table~\ref{tab:hist}).

Although the overall character of  the orbital motion is well defined,
the range  of inner  orbits compatible with  the data is  quite large,
more than suggested by the  formal errors of the elements.  Additional
consideration helps to select the most plausible inner orbit. Owing to
the faintness of Ab, it is safe to assume that the photo-center of the
composite  component  A   is  located  at  the  primary   Aa,  so  the
measurements of  A,B and Aa,B are  equivalent. In such  case, the mass
ratio in the inner pair $q_2  = {\cal M}_{\rm Ab}/ {\cal M}_{\rm Aa} =
1/(F -1)$.  The mass sum  of the inner  pair is proportional  to $(a_2
F)^3  P_2^{-2}$, the mass  sum of  the outer  pair is  proportional to
$a_1^3 P_1^{-2}$. As the stars  Aa and B have very similar brightness,
we expect equal masses  ${\cal M}_{\rm Aa} \approx {\cal  M}_{\rm B}$. So, the
ratio of the mass sums in the inner  and outer orbits should  be close to
$(1 + q_2)/(2 + q_2)$. This leads to
\begin{equation}
\frac{(a_2 F)^3 P_2^{-2}} {a_1^3 P_1^{-2}} \approx \frac{1 +
  q_2}{2 + q_2} .
\label{eq:masses}
\end{equation}
Both periods and the  outer semimajor axis $a_1$ are rather well defined,
hence the  inner semimajor  axis  should be  $F  a_2 \approx  0\farcs39$
according  to  this  equation.   The unconstrained  least-squares  fit
produces an inner  orbit with $e_2 = 0.96$ and  $a_2 = 0\farcs22$, too
large  to  match  the   expected  inner  mass  sum.   Convergence  of
astrometric  orbits  derived  from  noisy  data  to  nearly  parabolic
solutions has  been studied  by \citet{Lucy2014}, and  here we  have a
similar  situation.   As  the  line  of apsides  of  Aa,Ab  is  almost
perpendicular to the line of nodes ($\omega_2 \sim 110$\degr), increasing
the inclination $i_2$, the eccentricity $e_2$, and  the semimajor axis $a_2$ simultaneously has
little effect on  the apparent orbital ellipse. The  measures alone do
not constrain well enough the semimajor axis of the inner orbit and allow
solutions with unrealistically large $a_2$.

\begin{figure}
\epsscale{1.0}
\plotone{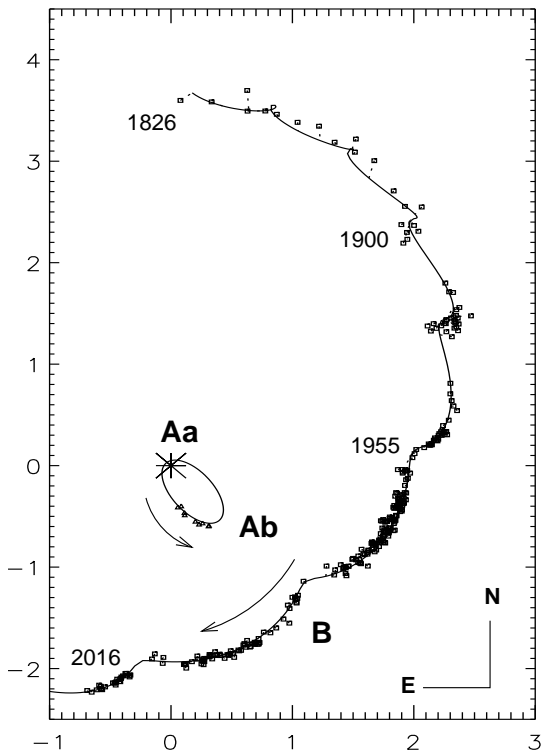}
\caption{Orbits of A,B  and Aa,Ab.  The component Aa  is at the
  coordinate origin. The curves show motions  of B and Ab around Aa on
  the  same scale (in  arcseconds); squares  and triangles  depict 
  observations of A,B and Aa,Ab, respectively.
\label{fig:AB}
}
\end{figure}

The final  orbits (Figure~\ref{fig:AB}) were obtained  by fixing $a_2$
to the value given by (\ref{eq:masses}) and fitting all other elements
plus  $F$.  Then  the  new value  of  $a_2$ was  computed and  the
constrained fit was repeated, until the iterations converged to $a_2 =
0\farcs110$.  There  is no  doubt  that the  inner  orbit  has a  high
eccentricity; for example, a fixed  $e_2 = 0.8$ leads to substantially
larger residuals.

Visual  measures in the  19th century  seem to  deviate systematically
from the orbit of A,B.  This  could be caused by an over-estimation of
the angular  separation. However, such residuals  might also result from
dynamical    interaction    between    the   orbits,    according    to
\citet{Harrington1968}.

One cannot help noting that the  outer and inner pairs move in the opposite
directions,  retrograde  and  prograde,  hence their  orbital  angular
momenta cannot be aligned.  There  are no radial velocity (RV) data of
adequate   precision   to  establish   the   correct  orbital   nodes.
Examination of the disparate published RVs indicates that  RV(A) was less
than RV(B)  in the first  half of the  20th century, meaning  that the
node $\omega_1=269\degr$ corresponds to the component A. The amplitude
of the RV difference between A and B is about 4\,km~s$^{-1}$, presently it is
+3 km~s$^{-1}$  according to the new  orbit.  However, the orbit  of Aa,Ab is
seen  almost face-on,  so the  RVs of  A might  be of  little help for
defining the inner node.  Fortunately, the two possible angles between
the orbital angular momentum vectors computed without knowledge of the
true nodes are close to each  other: $\Phi_1 = 146\fdg5 \pm 9\fdg9$ and
$\Phi_2 = 136\fdg9 \pm 10\fdg0$. The period ratio is $20.8 \pm 0.6$.

Relative  orientation of the  orbits and the  high inner  eccentricity are
strongly suggestive  of the Kozai-Lidov cycles in  this triple system.
If the inner  orbit were originally almost perpendicular  to the outer
orbit,  it would  evolve towards  high  eccentricity and
relative inclination  of 39\degr  or 141\degr.  In  a system  of three
 point masses the inclination  and inner eccentricity oscillate with
a period on the order of  $P_1^2/P_2 \sim 12$\,kyr, but real stars can
tidally  interact  at  periastron,  locking  the inner  orbit  in  the
high-$e$ state, with subsequent circularization \citep{KCTF}. However,
the distance  between Aa and Ab  at periastron is about  1.4 A.U., too
large for tidal interaction.

\begin{deluxetable}{r rrr rr }
\tabletypesize{\scriptsize}
\tablewidth{0pt}
\tablecaption{Observations and residuals of A,B (Fragment) \label{tab:obsAB}}
\tablehead{
\colhead{$T$} &
\colhead{$\theta$} & 
\colhead{$\rho$} &
\colhead{$\sigma$} & 
\colhead{O$-$C$_\theta$} & 
\colhead{O$-$C$_\rho$} \\
\colhead{(yr)} & 
\colhead{(\degr)} &
\colhead{($''$)} & 
\colhead{($''$)} & 
\colhead{(\degr)} &
\colhead{($''$)} 
}
\startdata
 1826.4000 & 358.8 & 3.6000 & 0.100 &  1.5 & $-$0.0793 \\
 1832.6000 & 354.7 & 3.6000 & 0.100 &  1.5 & 0.0295 \\
 1838.8900 & 350.4 & 3.7500 & 0.100 &  0.8 & 0.1918 \\
  \ldots  &  \ldots & \ldots & \ldots & \ldots & \ldots \\
 2014.7631 & 164.5 & 2.2928 & 0.009 &  0.0 & $-$0.0020 \\
 2015.7379 & 163.7 & 2.3261 & 0.003 &  0.1 & 0.0094 \\
 2016.3901 & 162.7 & 2.3212 & 0.003 & $-$0.2 & $-$0.0091 
\enddata
\end{deluxetable}

\begin{deluxetable}{r rrr rr }
\tabletypesize{\scriptsize}
\tablewidth{0pt}
\tablecaption{Observations and residuals of A{\rm a},A{\rm b} \label{tab:obsAa}}
\tablehead{
\colhead{$T$} &
\colhead{$\theta$} & 
\colhead{$\rho$} &
\colhead{$\sigma$} & 
\colhead{O$-$C$_\theta$} & 
\colhead{O$-$C$_\rho$} \\
\colhead{(yr)} & 
\colhead{(\degr)} &
\colhead{($''$)} & 
\colhead{($''$)} & 
\colhead{(\degr)} &
\colhead{($''$)} 
}
\startdata
 2009.7552 & 192.0 & 0.4057 & 0.009 &  4.9 & 0.0015 \\
 2009.7552 & 187.9 & 0.4115 & 0.009 &  0.8 & 0.0073 \\
 2010.8914 & 193.4 & 0.4760 & 0.009 &  0.5 & $-$0.0031 \\
 2010.9681 & 193.2 & 0.4976 & 0.009 &  0.0 & 0.0140 \\
 2012.9229 & 200.2 & 0.5835 & 0.009 &  0.2 & 0.0042 \\
 2013.7364 & 201.8 & 0.6211 & 0.009 & $-$0.4 & 0.0113 \\
 2014.7631 & 204.7 & 0.6218 & 0.009 & $-$0.1 & $-$0.0198 \\
 2015.7379 & 207.5 & 0.6700 & 0.009 &  0.4 & 0.0046 \\
 2016.3901 & 208.0 & 0.6665 & 0.009 & $-$0.5 & $-$0.0116 
\enddata
\end{deluxetable}

\begin{deluxetable*}{l l rrr rrr r ccc}
\tabletypesize{\scriptsize}
\tablewidth{0pt}
\tablecaption{Orbital Elements \label{tab:orb}}
\tablehead{
\colhead{System} &
\colhead{$P$} & 
\colhead{$T_0$} &
\colhead{$e$} & 
\colhead{$a$} & 
\colhead{$\Omega$} &
\colhead{$\omega$} &
\colhead{$i$}  &
\colhead{$F$}  \\
&   
\colhead{(yr)} & 
\colhead{(yr)} &
\colhead{ } & 
\colhead{($''$)} & 
\colhead{(\degr)} &
\colhead{(\degr)} &
\colhead{(\degr)} &
}
\startdata
A,B & 540        & 1981.50       & 0.419  & 3.496    & 131.3     & 269.3    & 142.0  & \ldots \\
      &$\pm$15  &$\pm$0.58  &$\pm$0.011   &$\pm$0.046 &$\pm$0.8  &$\pm$1.7  &$\pm$0.4 & \ldots\\ 
Aa,Ab& 25.95       & 2006.52     & 0.872  & 0.110     & 293.7   & 100.9         & 11.8  &  3.50  \\          
      &$\pm$0.048  &$\pm$0.13  &$\pm$0.006  & fixed  &$\pm$74  &$\pm$73  &$\pm$6.7 & $\pm$0.09 
\enddata
\end{deluxetable*}

\section{Physical parameters of the stars}
\label{sec:par}

\begin{figure}
\epsscale{1.0}
\plotone{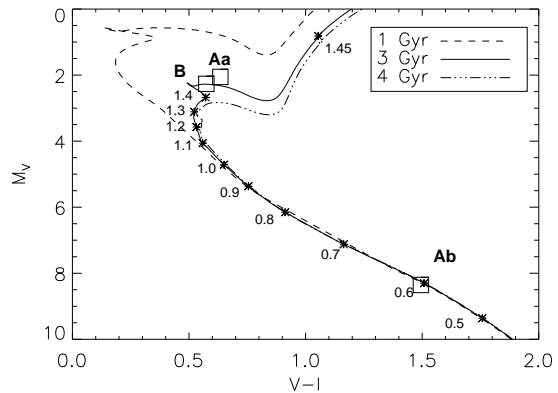}
\caption{Components of $\zeta$~Aqr on the $(M_V, V-I_C)$ color-magnitude
  diagram and Dartmouth isochrones for solar metallicity. Asterisks
  and small numbers show masses on the 3-Gyr isochrone.
\label{fig:iso}
}
\end{figure}

\begin{deluxetable}{l   ccc }[ht]
\tabletypesize{\scriptsize}
\tablewidth{0pt}
\tablecaption{Photometry \label{tab:ptm}}
\tablehead{
\colhead{Comp.} &
\colhead{$V$} & 
\colhead{$I_C$} &
\colhead{$K_s$}
}
\startdata
A+B     & 3.65 & 3.04 & 2.72 \\
B$-$A   & 0.21 & 0.27 & 0.33 \\ 
Ab$-$Aa & (6.3) & 5.44 & (4.0) \\
Aa      & 4.30  & 3.67 & 3.21 \\
B       & 4.51  & 3.94 & 3.93 \\
Ab      & (10.60) & 9.11 & (7.3)
\enddata
\end{deluxetable}

The {\it Hipparcos} parallax  of 35.5$\pm$1.3\,mas (distance modulus 2.25 mag)
relates orbital elements to masses.  The mass sum of AB is 3.3$\pm$0.3
${\cal  M}_\odot$  and  the  mass  sum  of  A  is  2.0$\pm$0.2  ${\cal
  M}_\odot$.  As noted  above, the inner mass sum is not tightly constrained,
 for this reason the near-equality of the masses of Aa and B was
imposed in  the orbit fit.  The masses  of Aa, B, and  Ab deduced from
the  orbits and  the mass  ratio  $q_2$ are  1.4, 1.4  and 0.6  ${\cal
  M}_\odot$, with an uncertainty of 10\%.

Available  combined  and   differential  photometry  is  collected  in
Table~\ref{tab:ptm}.  The  component A is brighter and  redder than B,
as the magnitude difference between A and B increases with wavelength:
$\Delta B = 0.14$ mag,  $\Delta V = 0.21$ mag \citep{Fab2000}, $\Delta
I   =  0.27$   mag   \citep{Hrt2012},   $  \Delta   K   =  0.33$   mag
\citep{Carbillet1996}.   Six measurements  of the  magnitude difference
between Aa  and Ab in the  $I$ band made  at SOAR average at  5.44 mag
with the rms scatter of 0.13 mag.  Considering that the absolute $I_C$
magnitude  of Ab  $M_I =  6.9$ mag matches  a main-sequence  star  of 0.6
${\cal  M}_\odot$,  its  magnitudes in $V$ and $K_s$  are  estimated  from  the
isochrone,  not measured  directly (numbers  in  brackets). Individual
magnitudes  of  the components  are  computed  from  the combined  and
differential photometry.

Figure~\ref{fig:iso}  places  the  components  of $\zeta$~Aqr  on  the
$(M_V, V-I_C)$ color-magnitude diagram.  The most massive stars Aa and
B are  located above the main  sequence and match  the 3-Gyr Dartmouth
isochrone for solar metallicity \citep{Dotter2008}.  The corresponding
masses  are 1.42  ${\cal M}_\odot$,  in excellent  agreement  with the
orbit. This means  that this system is unlikely  to contain additional
close  (spectroscopic) stellar companions  to any  of the  stars.  The
near-equality of the masses of A and B, adopted above as a hypothesis,
is supported by the isochrones  because at this evolutionary stage the
dependence of  color on  mass is  strong.  The age  of this  system is
close to 3\,Gyr.

The stars Aa and B have a fast axial rotation (about 50\,km~s$^{-1}$),
not  uncommon  for their  spectral  type  and  evolutionary status  as
subgiants.   Fast  rotation  is  related  to  the  high  chromoshheric
activity      and     X-ray      luminosity     \citep{Schroeder2009}.
\citet{Liebre1999}   measured  a   high  lithium   abundance   in  the
photosphere  of these  stars.   Most spectroscopic  studies treat  the
object as  a single star, given  that the light of  two closely spaced
and nearly  equal components Aa  and B is  usually mixed in  the slit.
When the spectra of both  components were taken separately, the unusual
slit illumination possiblly caused  discordant RVs and false claims of
RV variability. \citet{N04} found  a constant RV of 25.90\,km~s$^{-1}$
over  a 3.8-yr time  span in  the combined  light of  both components.
Using  this  RV  and  the  {\it Hipparcos}  astrometry,  the  Galactic
velocity  is  $(U,V,W)  =  (-16.0,+15.9,-28.3)$\,km~s$^{-1}$  ($U$  is
directed away from  the Galactic center). The spatial  motion does not
match  any  known  kinematic  group,  supporting the  view  that  this
multiple system is not very young.

\section{Discussion}
\label{sec:disc}

Comparable separations  in the inner  and outer orbits  of $\zeta$~Aqr
raise  concern  of  its  dynamical  stability.   There  are  no  exact
formulations for a criterion of dynamical stability of triple systems,
while the  existing approximate  criteria do not  differ substantially
between  themselves. One  of the  popular criteria  by  \citet{MA} for
coplanar orbits can be written as
%
\begin{eqnarray}
(P_{1}/P_{2})^{2/3} & \ge & 2.8 (1 + q_{1})^{1/15} \nonumber \\
           & \times & (1 + e_{1})^{0.4} (1 - e_{1})^{1.2} 
\label{eq:MA}
\end{eqnarray}
in  present notation. This  translates to  $P_1/P_2 >  16.2$ for  $q_1 =
0.70$ and $e_1 = 0.42$. According to this criterion, $\zeta$~Aqr with
$P_1/P_2 = 20.8$ is within  the stability boundary. On the other hand,
the   ``empirical stability criterion'' by \citet{Tok2004},
$P_{1}(1-e_{1})^{3}/P_{2}>5$, calls for $P_1/P_2 > 25.5$, placing this
triple system in  the instability zone. Given the  age of this system,
we know that it is in fact stable.  In any case, the dynamical interaction
between inner and  outer orbits should be strong,  and the description
of the motion by two Keplerian  orbits can certainly be improved by taking it
into account.

\begin{figure}
\epsscale{1.0}
\plotone{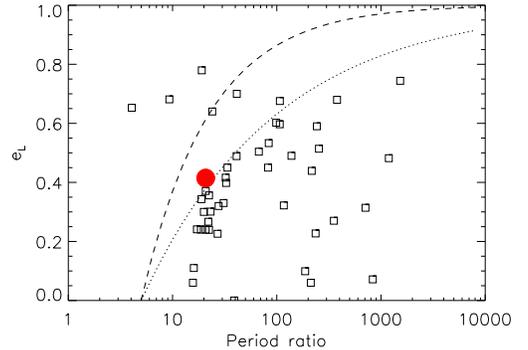}
\caption{Outer eccentricity $e_{\rm L}$  vs. period ratio $P_1/P_2$ in
  triple stars with  outer and inner orbits listed  in the Sixth Orbit
  Catalog (squares),  with the position of  $\zeta$~Aqr marked by  the red
  circle.  The dashed  and dotted  lines correspond  to  the dynamical
  stability criteria by \citet{MA} and \citet{Tok2004}, respectively.
\label{fig:stab} }
\medskip
\end{figure}

The stability  criteria can be  checked against multiple  systems with
known outer  and inner  orbits. A selection  of such systems  from the
current (2016)  version of  the Sixth Catalog  of Visual  Binary Stars
\citep{VB6}  is  shown in  Figure~\ref{fig:stab}.   Points above  both
curves are due to the  wrong orbital elements (visual orbits with long
periods  and small  coverage lack  credibility).  Even  such  a famous
binary as $\zeta$~Aqr  had outer orbits with periods  ranging within a
factor  of two (Table~\ref{tab:hist}).   There is  a group  of several
other triples  with similar outer eccentricities and  period ratios on
the  order of  20.   Three triple  systems  in this  group have  inner
secondary components of very low mass discovered by \citet{Mut2010}.

Properties of  multiple stellar systems  are related to  their origin.
The  orbital  architecture of  $\zeta$~Aqr  is  suggestive of  chaotic
dynamical interactions, rather than dissipative evolution in accretion
disks  associated   with  coplanar  multiple   systems  like  HD~91962
\citep{Tok2015}.  However, the masses of  the main components Aa and B
and their axial  rotation are remarkably similar (a  twin binary).  It
is likely that the axes of stars  Aa and B are aligned with each other
and, possibly, with the outer orbit, rather than with the inner orbit.
The  outer binary  Aa,B could  have  been formed  by a  collapse of  a
rotating  core with  subsequent  accretion that  made the  components'
masses nearly  equal. In this case,  the third star Ab  is an intruder
that met the binary Aa,B at a  later time and was captured on a nearly
perpendicular orbit  that is now  undergoing Kozai-Lidov oscillations.
The dynamical  capture of Ab could  happen in the  nascent cluster from
which this system  formed, or in an unstable  quadruple system where a
pair of  low-mass stars was  disrupted, leaving one of its components
bound to A,B and ejecting another. 
Some other  multiple systems with  a high mutual inclination  of their
orbits could be produced by such dynamical interplay.  \citet{Hei1984}
noted  the similarity  between  $\zeta$~Aqr and  the nearby  quadruple
system $\xi$~UMa  (ADS~8119).  The  latter consists of  two solar-mass
stars  on a  60-year outer  orbit  \citep{Hei1996}.  Each  star has  a
low-mass close companion.  The orbit of Aa,Ab is inclined to the outer
orbit A,B by  130\degr ~and has a substantial  eccentricity, just like
Aa,Ab in $\zeta$~Aqr.  The subsystem Ba,Bb with a period of 4 days and
a minimum Bb mass of only  0.04 ${\cal M}_\odot$ could be a product of
Kozai-Lidov cycles  with tidal dissipation  if its initial  orbit were
nearly  perpendicular  to the  orbit  of  A,B.   However, the  orbital
periods in $\xi$~UMa are $\sim$10 times shorter than in $\zeta$~Aqr.

The next periastron passage in the inner subsystem of $\zeta$~Aqr will
occur in 2032.6. Knowing that  the orbit is very eccentric, this event
should be  observed by spectroscopy  and imaging.  RV  variations near
the  periastron,  if detected,  will  inform  us  on the  exact  inner
eccentricity and  inclination.  Meanwhile, the motion  of Aa,Ab should
be monitored  by speckle interferometry in the  visible, while imaging
in the  near-infrared is needed to secure  differential photometry and
to measure  the colors of Ab.   The pair Aa,Ab is  now approaching its
maximum  separation  of  almost  0\farcs7, facilitating  its  resolved
photometry and, possibly, spectroscopy.

\acknowledgements

I thank B.~Mason  for extracting all measures and  references from the
WDS database.   This work used  the SIMBAD service operated  by Centre
des   Donn\'ees   Stellaires   (Strasbourg,   France),   bibliographic
references from  the Astrophysics Data System  maintained by SAO/NASA,
and the Washington Double Star Catalog maintained at USNO.


{\it Facilities:} \facility{SOAR}




\end{document}